# Electron transport properties of graphene nanoribbons with Gaussian deformation

**Van-Truong Tran**[1*], **Jérôme Saint-Martin**[2], **and Philippe Dollfus**[2]

[1]IMPMC, Université Pierre et Marie Curie (UPMC), Sorbonne Universités, 75252 Paris Cedex 05, France

[2]Université Paris-Saclay, CNRS, Centre de Nanosciences et de Nanotechnologies, 91120, Palaiseau, France.

Email: [*]vantruongtran.nanophys@gmail.com

## Abstract

Gaussian deformation in graphene structures exhibits an interesting effect in which flower-shaped confinement states are observed in the deformed region [Carrillo-Bastos *et al.*, Phys. Rev. B **90** 041411 (2014)]. To exploit such a deformation for various applications, tunable electronic features including a bandgap opening for semi-metallic structures are expected. Besides, the effects of disorders and external excitations also need to be considered. In this work, we present a systematic study on quantum transport of graphene ribbons with Gaussian deformation. Different levels of deformation are explored to find a universal behavior of the electron transmission. Using a tight-binding model in combination with Non-Equilibrium Green's Functions formalism, we show that Gaussian deformation influences strongly the electronic properties of ribbons in which the electron transmission decreases remarkably in high energy regions even if small deformations are considered. Interestingly, it unveils that the first plateau of the transmission of semi-metallic armchair ribbons is just weakly affected in the case of small deformations. However, significant large Gaussian bumps can induce a strong drop of this plateau and a transport gap is formed. The transmission at the zero energy is found to decrease exponentially with increasing the size of the Gaussian bump. Moreover, the gap of semi-conducting ribbons is enlarged with large deformations. The opening or the widening of the transport gap in large deformed armchair structures is interpreted by a formation of a three-zone behavior along the transport direction of the hopping profile. On the other hand, a transport gap is not observed in zigzag ribbons regardless of the size of Gaussian bumps. This behavior is due to the strong localization of edge states at the energy point $E = 0$. Furthermore, under the effect of a positive vertical electric field $+E_z$, it shows an enhancement of electron transport in the conduction region and a suppression in the valence one. The effect of a negative field $-E_z$ is reverse. Additionally, it is also pointed out that the electronic behavior of a Gaussian deformed ribbon including edge roughness is dominated by the characteristics of the edge-roughness effect with strong Anderson-type localized states reflected by sharp peaks in the transmission profile.

## I. Introduction

Graphene has been recognized as a material for the future of electronics due to its exceptional electronic properties with the extremely high electron mobility and a form of a thin layer structure. All these intriguing features could lead to compact and efficient electronic devices[1]. However, 2D graphene is a semi-metallic material[2] and exhibits an almost zero-bandgap that limits its possible applications in electronics. Interestingly, it has been demonstrated that narrow ribbons of graphene possess a finite bandgap[3,4] and promise to be suitable for different applications such as transistors[5,6], thermoelectric generators[7–9].

To push graphene ribbons closer to real applications, further studies of more realistic structures of ribbons containing defects such as vacancies[9–12] and edge roughness[10,13–16] have been taken up. It has been shown that such defects strongly influence the natural electronic properties of ribbons. Defects lead to a suppression of the electrical conductance depending on the vacancy position[9] and the level of vacancies[11]. Besides, edge roughness may lead to strong Anderson localization in areas of edge roughness[14].

Another type of disorder has been also paid attention which is deformation.

In-plane deformation in ribbon structures was first studied by Chang *et al*. in 2007[17] and, subsequently, additional works[18–21] have been carried out by other groups to provide a more comprehensive understanding of this effect. It has been unveiled that under a uniaxial strain, the electronic properties of zigzag graphene nanoribbons (ZGNRs) are almost unchanged while the bandgaps of armchair graphene nanoribbons (AGNRs) are observed to fluctuate with the applied uniaxial strain[18]. Grain boundary can also be considered as a local in-plane deformation[22,23] with a significant impact on the electronic transport properties[23].

On the other hand, the presence of out-of-plane deformation has been evidenced in many structures[24]. This kind of disorder has been examined recently on both 2D graphene[24–28] and ribbons [29–31]. It has been shown that graphene deposited on a low-quality substrate can contain out-of-plane deformations, and Gaussian deformations are frequently observed[24]. A Gaussian bump can also be generated during a Scanning Tunneling Microscope (STM) process when the STM tip can interact with the graphene layer via Van der Waals interactions[32].

The presence of deformation leads to a change in the mechanical properties of ribbons and also in the electronic ones. Recent studies have shown an interesting phenomenon in which flower-shaped confinement states are observed in centro-symmetric Gaussian deformed regions[30]. Additionally, a local sub-lattice breaking symmetry is found, i.e., an unequal distribution of charge density between the two nonequivalent sub-lattices A and B in the deformed region has been observed even for small deformations[25,33]. Valley-electronic filtering depending on geometrical deformation[34] and current-flow paths in the deformed region[26] have also been discussed for 2D graphene with Gaussian deformation.

Although several works have been carried out to unveil the changes in the electronic properties of graphene in the presence of Gaussian deformation, the number of studies in this topic remains modest. In particular, previous studies of quantum effects in Gaussian-deformed ribbon structures have been still limited to consideration of confinement states[30,31] and charge distribution at sub-lattice sites[30]. Moreover, only Gaussian bumps with the size smaller than the width of considered ribbons have been investigated. Thus, further studies are needed to understand more comprehensively the impact of the shape of Gaussian bumps on the electronic properties according to the size of ribbons. Furthermore, the impacts on electron transport in a Gaussian deformed ribbon of an external electric field and edge roughness have not yet been considered.

In this work, we aim at investigating systematically the impact of Gaussian deformation on electron transport in both armchair and zigzag graphene ribbons. The correlation between the shape of a Gaussian bump and the size of a studied ribbon will be explored. In particular, we pay attention to energy gap opening in semi-metallic ribbons in order to optimize graphene-based atomistic designs suitable for a broad range of applications. In addition, the effects of an external electric field and edge roughness are also considered.

The rest of the paper is organized as follows: in Sec. II we first present the concept of Gaussian deformation in a ribbon structure and the parameters used to define the Gaussian shape and the size of a ribbon, then a tight-binding (TB) model and Non-Equilibrium Green's Functions (NEGF) formalism are detailed for methodology. Sec. III is devoted to results and discussions. In Sec. III A, a comprehensive study of the electron transport in graphene ribbons with Gaussian deformation is discussed. In Sec. III B, the effect of an external electric field on the electronic properties of Gaussian deformed ribbons is investigated. In Sec. III C, the individual and mutual impacts of Gaussian deformation and edge roughness are presented. Finally, Sec. V concludes this paper.

## II. Model and methodology

### A. Model

A graphene ribbon with Gaussian deformation is illustrated in Fig. 1 for two typical sizes of Gaussian bumps: small (Fig. 1(a)) and large (Fig. 1(b)), with respect to the size of the ribbon. The bump position is illustrated with a color gradient. These two distinct Gaussian bumps may impact differently the physical properties of the ribbon including the electron transport properties.

The height of atoms within a centro-symmetric Gaussian deformed region is defined as follows[35]

$$z(x,y) = h_G e^{-\frac{(x-x_0)^2 + (y-y_0)^2}{2\sigma^2}}, \qquad (1)$$

where $x_0$, $y_0$ are the $x$, $y$ coordinates of the central point of the bump. In all cases, we set the peak of the bump to be on top of the center of the considered ribbon. The shape of the Gaussian bump is generally characterized by two geometrical parameters $h_G$ and $\sigma$ which are the height and the standard deviation of the Gaussian shape as illustrated in Fig. 1(c)[35]. Sometimes, the parameter $b = \sqrt{2}\sigma$ is also used to characterize the width of a Gaussian bump[30,33,34]. To compare the width of the Gaussian bump with that of the ribbon, it may be more relevant to determine the diameter of the bottom circle of the Gaussian shape on the plane of the ribbon. It is well known that the circle of radius $R = 3b/\sqrt{2} = 3\sigma$ contains 99.7% of the Gaussian bump[31]. The diameter of this circle is thus $W_G = 2R = 6\sigma$ and can be used as the bump width to be compared with the width of the ribbon.

Besides, the geometrical definition of the width of a ribbon depends on the edge orientation of the ribbon. The width of an AGNR is defined as $W_R = (M-1) \times \sqrt{3}/2 \times a_0$, while that of a ZGNR is calculated as $W_R = (3M-2) \times a_0/2$, where $a_0 = 1.42$ Å is the distance between two neighboring atoms in perfect graphene structures and $M$ is the number of dimer (chain) lines along the ribbon width of an AGNR (ZGNR)[36].

It should be noted that in the area of a Gaussian bump, the mechanical strain is non-uniform and the strain intensity inside the deformed region is defined as $\varepsilon = h_G^2/(2\sigma^2)$[33]. On the other hand, it has been shown in previous studies[37] that for a strain intensity above 25%, graphene enters into the inelastic regime where both the mechanical and electronic properties are unpredictable. Therefore, in this work, we consider only Gaussian bumps with strain intensity less than or equal to 15%.

It is also worth mentioning that Gaussian bumps generated by rough substrates may be of a few nm size[24]. The diameter of an STM tip being in the range of a few nm to 50 nm[38–40], Gaussian bumps generated by an STM process can have a width ranging from less than 1 nm to a few tens of nm. So, in this work we consider only Gaussian deformation of such width, to be consistent with experiments. Additionally, ribbons having the width of a few nm are now feasible by means of the latest technology[41,42].

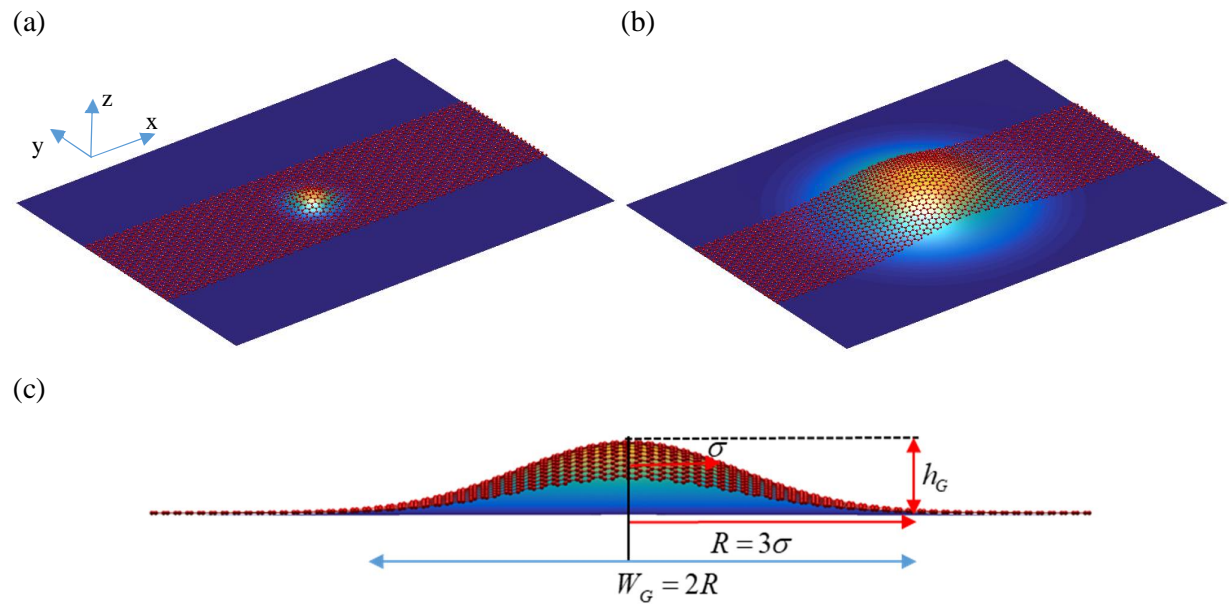

**Fig 1**: Sketch of a ribbon with Gaussian deformation. Two typical deformations are illustrated: (a) small and (b) large Gaussian bumps. (c) Parameters defining the shape of a Gaussian bump.

### B. Methodology

To investigate the electronic properties of non-uniform strain graphene structures, we employed the simple first nearest neighbors (1NN) TB model that has been widely used in previous works[26,30,31,33,35]. A term related to a non-zero external electric field was also added. Thus, our Hamiltonian yields:

$$H = -\sum_{ij \in 1NN} t_{ij} |i\rangle\langle j| + \sum_i U_i |i\rangle\langle i| \qquad (2)$$

The hoping energy between two lattice sites $i$-th and $j$-th is defined as

$$t_{ij} = t_0 e^{-\beta\left(\frac{d_{ij}}{a_0}-1\right)}, \qquad (3)$$

where the coefficient $\beta = 3.37$ is defined by the strain theory[43], $t_0 = 2.8$ eV is the hopping energy between the two nearest sites in the unstrained region, $d_{ij} = \sqrt{(x_i-x_j)^2+(y_i-y_j)^2+(z_i-z_j)^2}$ is the distance between the $i$-th and $j$-th sites. $U_i = -e.\vec{E}*(\vec{r_i}-\vec{r_O})$ is the electrostatic potential at the $i$-th lattice site under an external field $\vec{E}$ and $\vec{r_O}$ is the origin of the potential.

The quantum transport properties of structures were examined by coupling the TB Hamiltonian with the NEGF technique[44]. All structures were divided into three parts: the left and right leads and the device region (central region). The leads were considered as semi-infinite regions. The device (central) region contains the left lead extension, the active region, and the right lead extension and these parts have $N_L$, $N_A$, $N_R$ primary unit cells, respectively. The length of the device is characterized by the total number of unit cells $N = N_L + N_A + N_R$. Disorders were introduced only in the active region. It is worth noting that a primary unit cell of a ribbon contains two slices with a total of $2M$ atoms. In our calculations, $N_L$ and $N_R$ were chosen equal to 5 unit cells, which is sufficient to make the left (right) lead isolated from the active region.

The Green's function of the device region was calculated as follows

$$G = \left[E^+.I - H_D - \Sigma^s_L - \Sigma^s_R\right]^{-1}, \qquad (4)$$

where $E^+ = E + i.\eta$ with $\eta$ is an infinitesimal positive number added to the energy to avoid the possible divergence of Green's functions, $H_D$ is the Hamiltonian of the device and

$$\Sigma^s_L = \left(E^+.I - H_{DL}\right)G^0_L\left(E^+.I - H_{LD}\right)$$
$$\Sigma^s_R = \left(E^+.I - H_{DR}\right)G^0_R\left(E^+.I - H_{RD}\right) \qquad (5)$$
6

define the surface self-energies contributed from the left and right leads. $G^0_{L(R)}$ represents the surface Green's function of the isolated left (right) lead and was computed by Sancho's technique[45]. The size of the device Green's function in Eq. (4) was reduced using the recursive technique[46]. Then electron transmission was computed as[44,47]

$$T_e = Trace\left\{\Gamma^s_L\left[i\left(G_{11} - G_{11}^\dagger\right) - G_{11}\Gamma^s_L G_{11}^\dagger\right]\right\}, \quad (6)$$

where $\Gamma^s_{L(R)} = i\left(\Sigma^s_{L(R)} - \Sigma^{s\,\dagger}_{L(R)}\right)$ denotes the surface injection rate at the left (right) lead. The local density of states (LDOSs) at the $i$-th lattice site were calculated by[46]

$$D(\vec{r}_i, E) = -\frac{\text{Im}\left[G_{ii}(E)\right]}{\pi} \quad (7)$$

### III. Results and discussions

In this section, first the impact of Gaussian deformation on the electron transport properties of different groups of ribbons is analyzed in detail. Then the variation of the electronic properties of deformed ribbons under an external electric field and the presence of edge roughness is discussed.

#### A. Impact of Gaussian deformation on the electronic properties of ribbons

It is well known that based on the electronic features, perfect AGNRs are classified into three groups $M = 3p + 2$, $3p + 1$ and $3p$ with $p$ is an integer number[4]. Thus, to understand precisely the impact of Gaussian deformation on the electronic properties of different types of ribbons, it is necessary to examine the effect of deformation for each of these groups.

First, we investigate AGNRs of the semi-metallic group $M = 3p + 2$. In Fig. 2, the electron transmission of a device made of a semi-metallic AGNR of width $M = 41$ ($W_R \approx 49.19$ Å) and length $N = 150$ unit cells ($L \approx 637.58$ Å) is shown for the perfect (undeformed) structure (black curve) and for deformed ones with several configurations of Gaussian bumps: very small ($W_G / W_R \approx 0.67$, red curve), small ($W_G / W_R \approx 0.95$, violet curve), medium ($W_G / W_R \approx 1.50$, blue curve) and large ($W_G / W_R \approx 5.63$, green curve) bumps. In all cases, strain intensity is fixed at 15%. As can be observed, the electron transmission is altered even for small Gaussian bumps (red and violet curves) where $W_G < W_R$. The degradation is more pronounced in the high energy regions than in the low energy region. Notably, the first step of the transmission remains almost unchanged. The effect is stronger with larger bumps. This result is in agreement with what was observed in a previous study[30]. Interestingly, when Gaussian deformation is large enough (blue, green lines) with $W_G > W_R$, the transmission at high energy is found to weakly change. However, the first plateau drops strongly around $E = 0$ and a transport gap is formed when the Gaussian bump is sufficiently large (green line). At the energy point $E = 0$, the transmission of the strongly deformed ribbon remains at only 1.6% ($T_e = 0.016$) compared to the value of 1.0 in the perfect and weakly deformed structures.

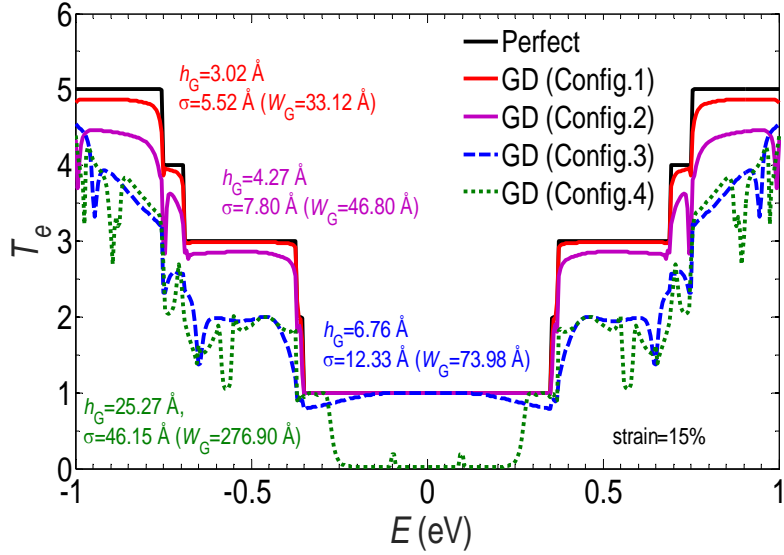

**Fig. 2**: Electron transmission in an AGNR without and with Gaussian deformation (GD): $M = 41$ ($W_R \approx 49.19$ Å), $N = 150$ ($L \approx 637.58$ Å). Different levels of deformation are considered and strain intensity is fixed at 15%.

To understand better the variation of the electron properties due to the Gaussian deformation, the LDOS is plotted in Fig. 3 as a function of energy and the transport direction (ox) in real space. Figs. 3(a) and 3(b) respectively present the results of the medium (Config. 3) and large (Config. 4) deformed structures shown in Fig. 2. In Fig. 3(b), i.e., in the case of large deformation, we observe that the LDOS near the peak of the Gaussian bump (central position) around the energy $E = 0$ is strongly reduced (dark blue area) with respect to the case of small deformation presented in Fig. 3(a) (blue area).

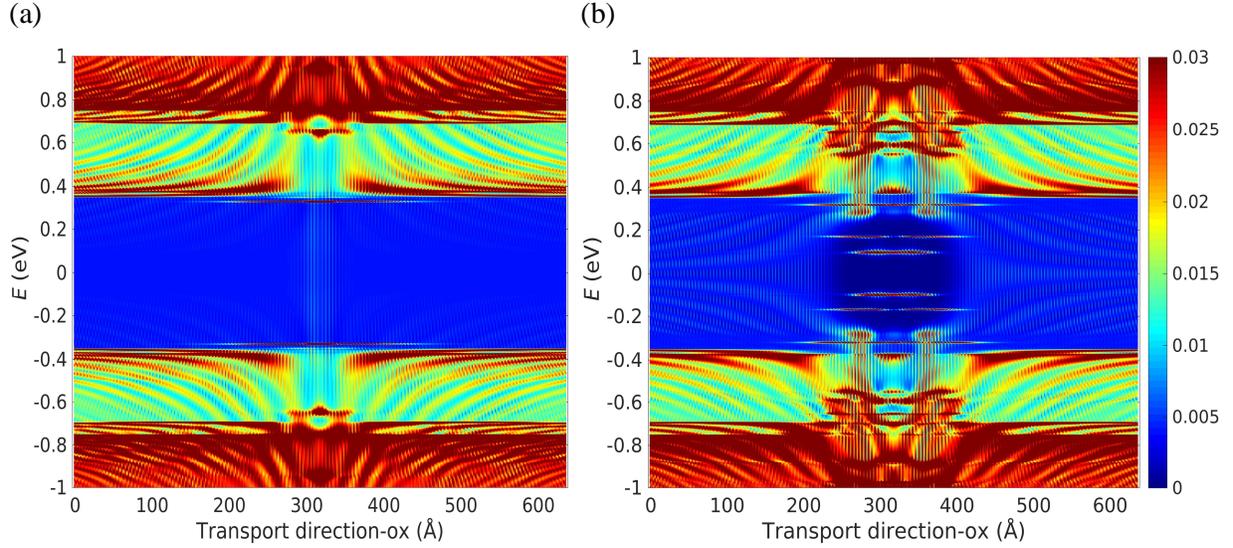

**Fig. 3**: LDOS (in arb. unit) as a function of energy and transport direction (ox) for two deformations: $h_G = 6.67$ Å, $\sigma = 12.33$ Å (Config. 3) and $h_G = 25.27$ Å, $\sigma = 46.15$ Å (Config. 4). $M = 41$ ($W_R \approx 49.19$ Å), $N = 150$ ($L \approx 637.58$ Å).

The low LDOS in the central region can be understood as a consequence of the larger distance between atoms in the deformed region. It induces smaller hopping energy than in the case of perfect or weakly-deformed structures and also smaller electrical conduction in the deformed region. This is consistent with the transmission drop around the zero-energy shown in Fig. 2.

To clarify this point, we studied the bonding lengths in each considered structure. The bonding maps for the centro-symmetric Gaussian deformations configs. 3 and 4 are respectively shown in Figs. 4(a) and 4(b). The color gradient presents the changes in nearest bonds (1NN) with atom distance ranging from 1.42 Å to 1.5 Å. From these panels, it manifests that the strongest deformation occurs at the middle height of the Gaussian bumps, while the areas at the top and the leg of the bumps are weakly tensile. It is worth noting that the profile of bonds in Fig. 4(a) leads to a six-folds region with low LDOSs which is similar to the flower shape observed in the previous study[30].

From the calculated bonding length, the nearest hopping energy of each bond was deduced by using Eq. (3). In Figs. 4(c) and 4(d) the profile of the hopping energies of all bonds at their bonding positions along the transport direction is shown for the two considered structures. The coordinates of a bond between the $i$-th and $j$-th atoms were simply defined as $\vec{r}_{t_{ij}} = \left(\vec{r}_i + \vec{r}_j\right)/2$. Interestingly, the shape of these two hopping profiles is remarkably different and can be used to interpret the physics involved in the behavior of the corresponding transmissions shown in Fig. 2. Fig. 4(c) indicates that in the weakly deformed structure, a single non-uniformed region is formed around the top of the Gaussian bump where the hopping energies are reduced. Similar hopping profiles are obtained for Configs. 1 and 2. Differently, in the case of strong deformation, three non-uniformed regions are formed inside the Gaussian bump as the standard hoping energy region near the peak is sandwiched in between two low hoping energy regions. The electron transport between regions of different hoping energies is limited and even blocked in the case of strong deformation. Indeed, in such a three-region system which is similar to a double-barrier potential profile, the presence of scatterings at each region interface and also quantum trapping in the pseudo "well" limit the electron transport. That explains the low transmission in the low energy region for the Config. 4 observed in Fig. 2. Moreover, the difference between the hopping profiles of Configs. 3 and 4 stems from the correlation between the shape of the Gaussian bump and the size of the ribbon. In Config. 3, the hopping profile along transport direction (Fig. 4(c)) shows that hopping energies in the region around the peak of the Gaussian bumps are similar to the those of the regions on the two sides of the peak and there is a significant increase of some bonding lengths along the y-direction in this region (Fig. 4(a)). In contrast, if the Gaussian bump along the y-direction covers the full width of the ribbon (Fig. 4(b)), the region around the peak of the large deformed structure contains only weakly stressed bonds and thus hopping energies are remarkably different from those in the two regions on the sides on the peak (Fig. 4(d)).

In summary, when the deformation is distributed over the entire width of the ribbon, the low energy electrons cannot cross the slightly deformed regions near the edges as in Config. 3 (see Fig. 4(a)). And thus a transport gap is observed when the Gaussian bump is large enough so that the hopping profile of the deformed region clearly shows a three-zone characteristic.

To further analyze the dependence of the transmission on the level of deformation, and particularly to find out the crucial condition to observe a transport gap around zero-energy point, we investigated the variation of the transmission at $E = 0$ for different configurations of Gaussian deformation. As the shape of a Gaussian bump depends not only on the height $h_G$ but also the width $\sigma$ or $W_G$, it is relevant to consider these two parameters. Additionally, the transport properties also depend on the ribbon width $W_R$, so it can be more relevant to consider the correlation between the shape of the Gaussian bump and that of the ribbon. In Fig. 5, $T_e(E$

= 0 ) is plotted as a function of the ratio $h_G W_G / W_R$ for different ribbon sizes and also different levels of strain induced by Gaussian deformation.

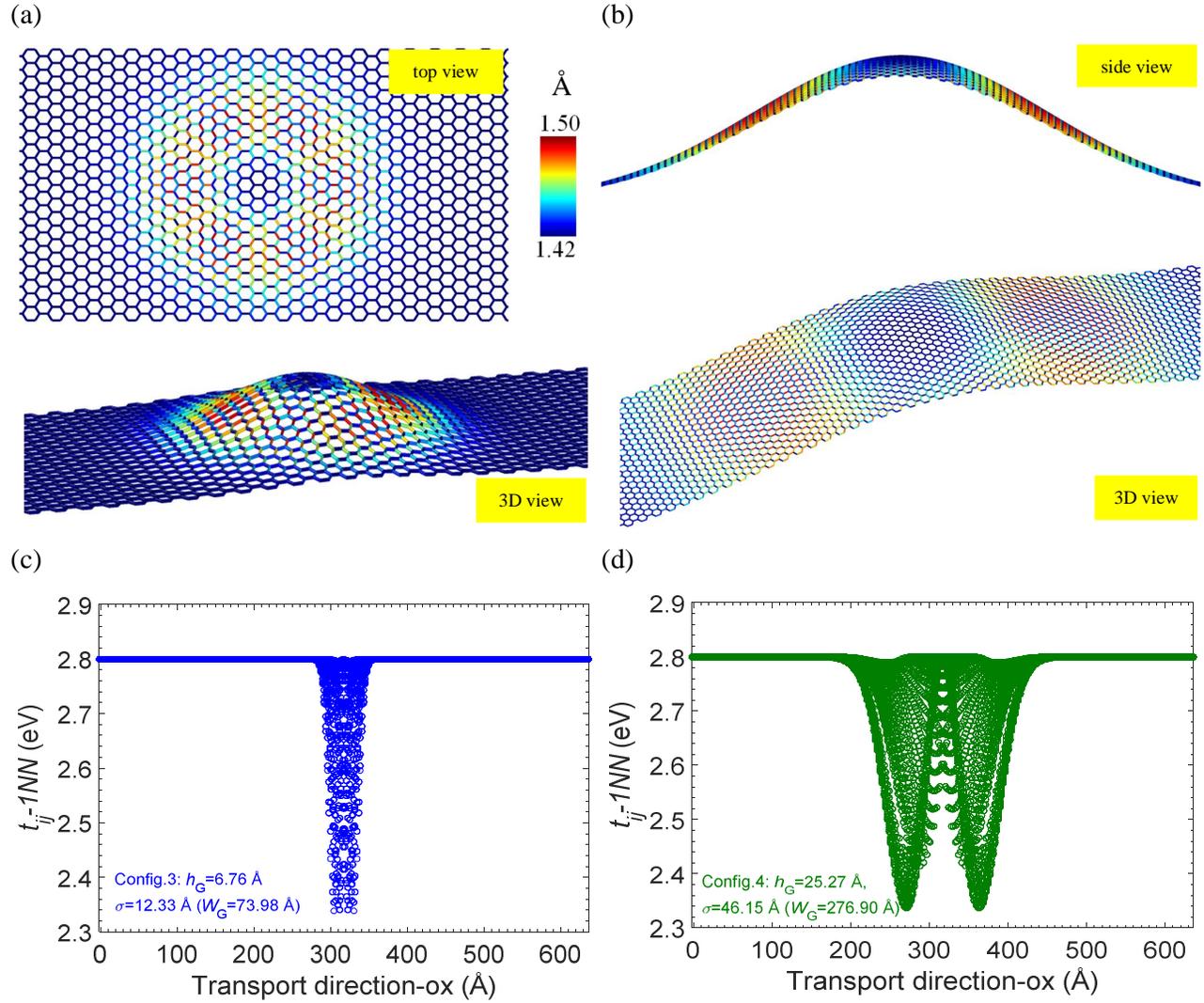

**Fig. 4**: Bonding length map and hopping profile in the ribbon with two different centro-symmetric Gaussian deformed configurations: (a) & (c) $h_G$ = 6.67 Å, $\sigma$ = 12.33 Å (Config. 3) and (b) & (d) $h_G$ = 25.27 Å, $\sigma$ = 46.15 Å (Config. 4). Here $M$ = 41 ($W_R \approx$ 49.19 Å), $N$ = 150 ($L \approx$ 637.58 Å).

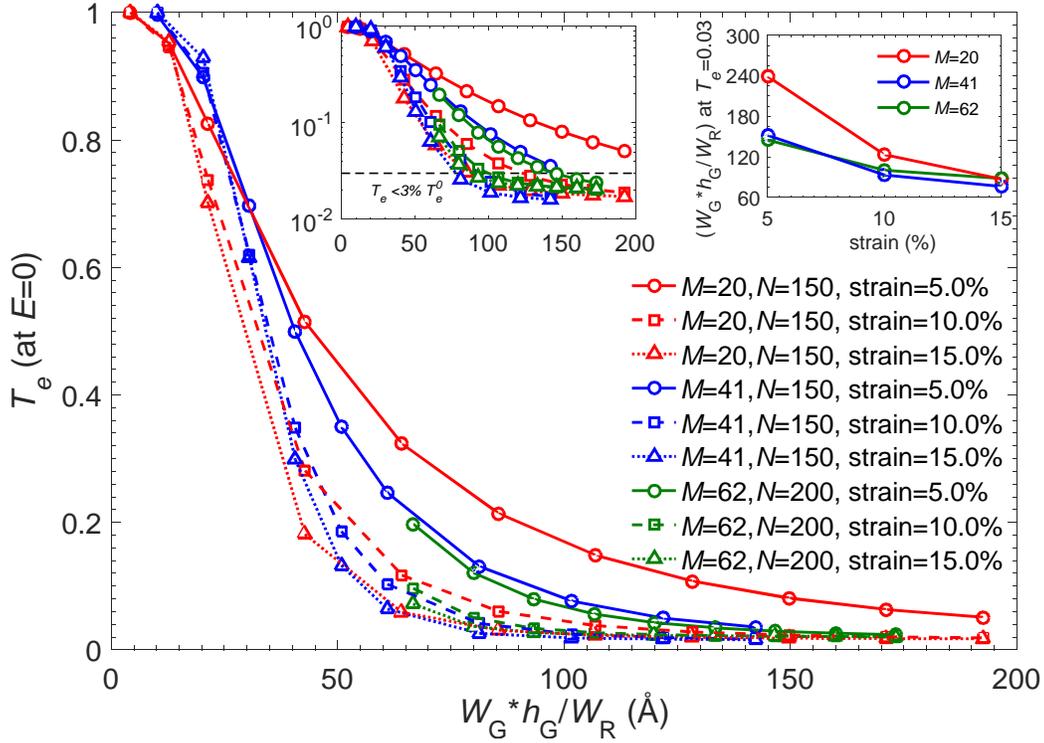

**Fig. 5**: Correlation between $T_e$ at $E = 0$ and the ratio $h_G W_G / W_R$. Left inset is the result that $T_e$ is in the logarithmic scale and the horizontal dash line in this inset indicates that 97% transmission is suppressed. Right inset shows the cut-off ratio $h_G W_G / W_R$ at $T_e = 0.03$ of each curve shown in the main panel.

As can be observed in the left inset of Fig. 5, all the transmission curves drop almost exponentially. Interestingly, these curves tend to converge at a large value of the ratio $h_G W_G / W_R$. It is shown in the left inset that below the horizontal dash line, the transmission is reduced by 97% confirming that an effective transport gap is formed when the shape of the Gaussian bump is large enough. For each configuration, the ratio $h_G W_G / W_R$ leading to this transmission reduction of 97% (i.e. at $T_e = 0.03*T_e^0$) was determined by performing a spline fitting. The results are shown in the right inset of Fig. 5. It can be observed that the ratio to reach the transmission threshold of $0.03*T_e^0$ depends on both the ribbon width and the level of strain while the latter is associated directly with the shape of the Gaussian bump. Besides, when ribbons are large enough and the strain is significant (equal to or larger than 15%), the required ratio seems to converge to a value of about 80.

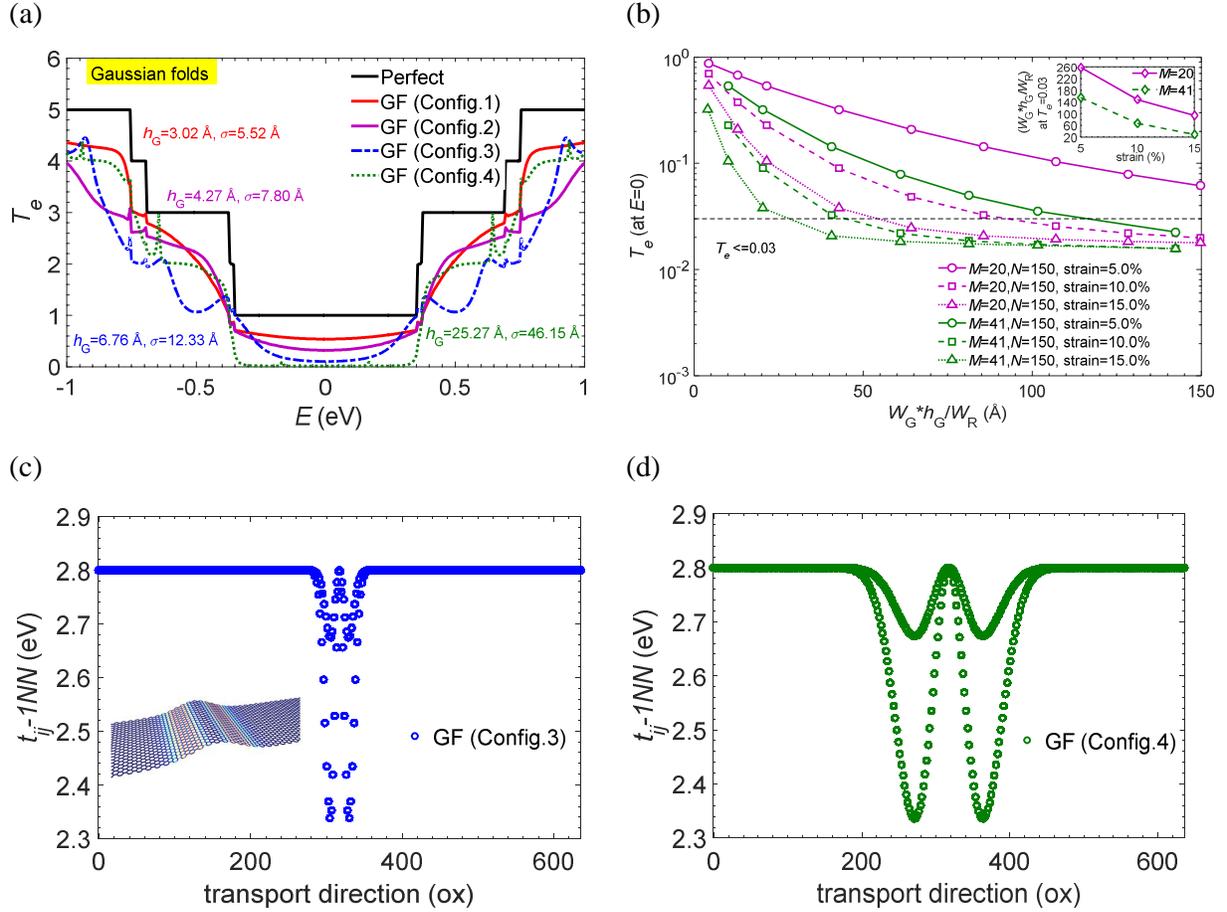

**Fig. 6**: Gaussian folds (GF): (a) $T_e$ as a function of energy for several GF configurations in the ribbon of width $M = 41$ ($W_R \approx 49.19$ Å), and length $N = 150$ ($L \approx 637.58$ Å). (b) Correlation between $T_e$ (at $E = 0$) and the ratio $h_G W_G / W_R$. (c) & (d) Hopping profiles in the ribbon with the GF Configs. 3 & 4 presented in Fig. 6(a). Inset in Fig. 6(b) presents the cut-off ratio $h_G W_G / W_R$ at $T_e = 0.03$ as a function of strain for different ribbon widths. Inset in Fig. 6(c) is the bonding map of the GF Config. 3, the color bar is from 1.42 Å to 1.5 Å as in Fig. 4.

It is worth noting that the large centro-symmetric Gaussian deformed structure (bump) shown in Fig. 4(b) is similar to a ribbon with a Gaussian fold where the height of an atom in the deformed region is defined as $z(x, y) = h_G e^{-\frac{(x-x_0)^2}{2\sigma^2}}$. This type of Gaussian deformation generates a fold along the y-direction. Such a deformation has been discussed recently about valley filtering properties[34] and Kondo effect under a magnetic impurity in 2D graphene structures[28]. To understand if the formation of a transport gap is also observed in graphene ribbons with this kind of Gaussian deformation, we examined the electronic properties of several Gaussian fold deformed ribbons. The obtained results are displayed in Fig. 6. Fig. 6(a) presents the transmission of a ribbon of width $M = 41$ ($W_R \approx 49.19$ Å) and length $N = 150$ ($L \approx 637.58$ Å) for four different Gaussian fold configurations as displayed in this panel. As can be observed in Fig. 6(a), with the same Gaussian shape parameters, the transmission of a Gaussian fold deformed structure is degraded more strongly compare to that of its Gaussian bump counterpart (shown in Fig. 2). This can be understood as the Gaussian fold with the same Gaussian parameters has a larger deformed surface compared to the similar Gaussian bump (see insets of Fig. 6(c) and Fig. 4(a)). Transmission at $E = 0$ is also found to drop exponentially as in the case of the Gaussian bumps as shown in Fig. 6(b). However, to achieve a transmission reduction of 97% in the case of Gaussian folds, weaker deformation than that in the case of bumps is required. Indeed, for instance in the case of 10% of strain and a ribbon of width $M = 41$, the crucial ratios $h_G W_G / W_R$ are equal to 67 for the fold and 93 for the bump. An analysis of the hopping profiles in these structures with Gaussian fold seen in Figs. 6(c) and 6(d) gives behavior similar to those related to the Gaussian bumps, i.e., a transport gap is observed when the hopping profile exhibits a three-zone characteristic. It should be noted that the hopping profile of a Gaussian fold ribbon is different from that of a Gaussian bump ribbon because at a given $x$ coordinate, the deformation along the y-direction is uniformed in this type of deformed structures as can be seen in the inset of Fig. 6(c).

Thus, qualitatively the impact of Gaussian bumps and folds on the electron transport properties of graphene ribbons is similar. Therefore, only deformations with Gaussian bumps are discussed further hereafter.

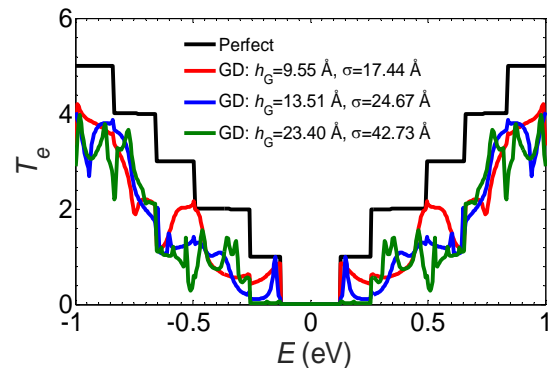

**Fig. 7**: Semiconducting Armchair with several configurations of centro-symmetric Gaussian deformation (GD): $T_e$ versus $E$. Here $M = 39$ ($W_R \approx 46.73$ Å), $N = 150$ ($L \approx 637.58$ Å).

To understand the effect of Gaussian deformation on the electronic properties of semiconducting AGNRs (groups $3p + 1$ and $3p$), we examined a ribbon of width $M = 39$ ($W_R \approx 46.73$ Å) which belongs to the $3p$ group. The results are shown in Fig. 7 for different shapes of Gaussian deformation (Gaussian bumps). Similar results were obtained for the $3p + 1$ group (not shown).

As can be seen in Fig. 7, around the first step of the "perfect" transmission (black line), in the presence of Gaussian deformation, transmission decreases and the reduction is stronger with larger Gaussian bumps. Interestingly, with the largest Gaussian deformation considered here, the bandgap seems to be enlarged (green line). For transmission at higher transmission steps, we also observe a strong reduction for all studied cases. Additionally, the behavior is similar to that of semi-metallic ribbons in which the high-energy transmission of the deformed structures is less dependent on the level of deformation. Our analyses showed that the variation of transmission at the low energy region around $E = 0$ is similar to that of the semi-metallic group ($M = 3p + 2$) as discussed above, i.e., the transport gap is only widened when the hopping profile presents three non-uniformed zones along the transport direction in the deformed region.

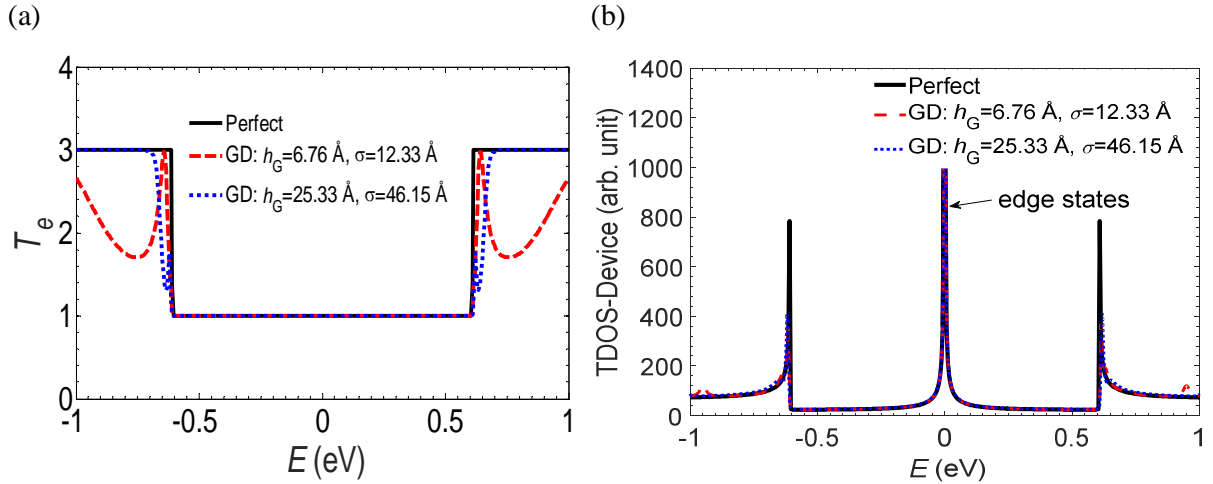

**Fig. 8**: (a) Transmission and (b) TDOS of a device made of a ZGNR of width $M = 20$ ($W_R \approx 41.18$ Å), $N = 150$ ($L \approx 367.69$ Å).

To complete the examination of the different types of ribbons, we now consider Gaussian deformation in ZGNRs. In Fig. 8, the results of the electron transmission of a zigzag structure of width $M = 20$ chain lines ($W_R \approx 41.18$ Å) are shown. Similar results were obtained for other ZGNRs. As can be seen in Fig. 8(a), the transmission without and with Gaussian deformation displays a reduction of the electron transport near the first step of the transmission. However, the first plateau remains unchanged even with a strong deformation (dot blue line). Thus there is no bandgap opening in ZGNRs under Gaussian deformation. This phenomenon can be understood by looking at the total density of states (TDOSs) in the device. The TDOSs without and with deformations are shown in Fig. 8(b). It can be seen that the middle peak of the TDOSs localizes at the energy $E = 0$ and it is unchanged with the deformation. This peak actually corresponds to the strongly localized edge states in the zigzag ribbon[3]. Due to the presence of edge states in the low energy region, even for strong deformations a transport gap cannot be opened.

**B. The asymmetrical effect of a vertical electric field on Gaussian deformed ribbons**

It has been demonstrated that external electric fields can be used to modulate the electronic properties of materials and devices[2,48,49]. It has been shown in previous studies[48,50,51] that a transverse (positive or negative) electric field modulates symmetrically the conduction and valence bands. This phenomenon stems from the mirror symmetry of ribbons about an axis located in the middle of ribbons. A similar effect has also been observed in bilayer structures with a vertical electric field[2,52].

In the presence of a Gaussian deformation, the mirror symmetry in the transverse plane (xy plane) is not affected but this symmetry in the vertical planes is broken. Thus some asymmetrical effects could be observed if a vertical electric field was applied.

To verify this prediction, we first examined the effect of a transverse electric field $\vec{E} = E_y.\vec{e_y}$ in a Gaussian deformed AGNR of width $M = 41$ (group $3p + 2$), length $N = 150$. For different external fields, the transmission is plotted as a function of the energy in Fig. 9(a). The results without any external field for the perfect and deformed structures are also displayed for comparison. As can be seen, the transverse electric field remarkably impacts on the electron transport, particularly in the low energy region around $E = 0$. This electric field acts symmetrically on the conduction and valence bands. Additionally, as observed in standard ribbons[53], the sign of the field is not relevant, only the norm of the field matters.

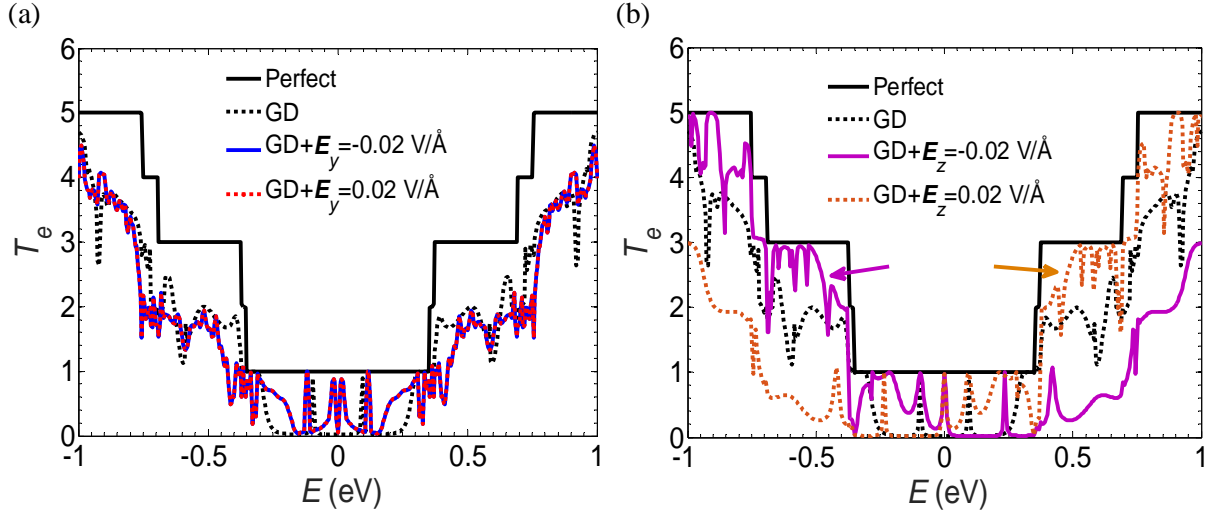

**Fig. 9**: An AGNR of width $M = 41$ ($W_R \approx 49.19$ Å) and length $N = 150$ ($L \approx 637.58$ Å) with a large Gaussian deformation (GD) $h_G = 25.27$ Å, $\sigma = 46.15$ Å under the effect of (a) a transverse electric field $\vec{E} = E_y.\vec{e_y}$ and (b) a vertical electric field $\vec{E} = E_z.\vec{e_z}$.

It is worth noting that within a more sophisticated TB model up to third nearest neighbors and with overlap factors, the conduction and valence bands are not perfectly symmetrical[36]. However, in TB models an external field modifies only the onsite energy of atoms and it induces an effect on both conduction and valence bands, thus the change attributed to the external fields does not depend on the chosen TB model. Also, as the studied device sizes are significantly large (from a few thousand to more than ten thousands of atoms), to avoid a computationally-demanding self-consistent process, fields lower than 20 mV/Å were considered and charge redistribution in such large ribbons was neglected.

To check the effect of a vertical electric field $\vec{E} = E_z.\vec{e_z}$, the transmissions computed for both +$E_z$ and -$E_z$ are shown in Fig. 9(b). Interestingly, as predicted, an asymmetrical effect on the conduction and valence bands is observed with both +$E_z$ (orange curve) and -$E_z$ (violet curve). Under a +$E_z$ field, the electron transport in the conduction range is significantly enhanced (compared to the case without electric fields) although it remains lower than the transmission of the perfect (undeformed) structure. In contrast, the transmission in the valence band is reduced. It is also worth noting that the transport gap is shifted to the position below the energy point $E = 0$. Furthermore, the effect of the -$E_z$ field is reverse, i.e., an enhancement of the transmission in the valence band and a reduction in the conduction band, as seen in Fig. 9(b).

This asymmetrical effect of the vertical electric field on the electron transmission may be interesting for several applications such as energy filters, rectification devices, or sensors.

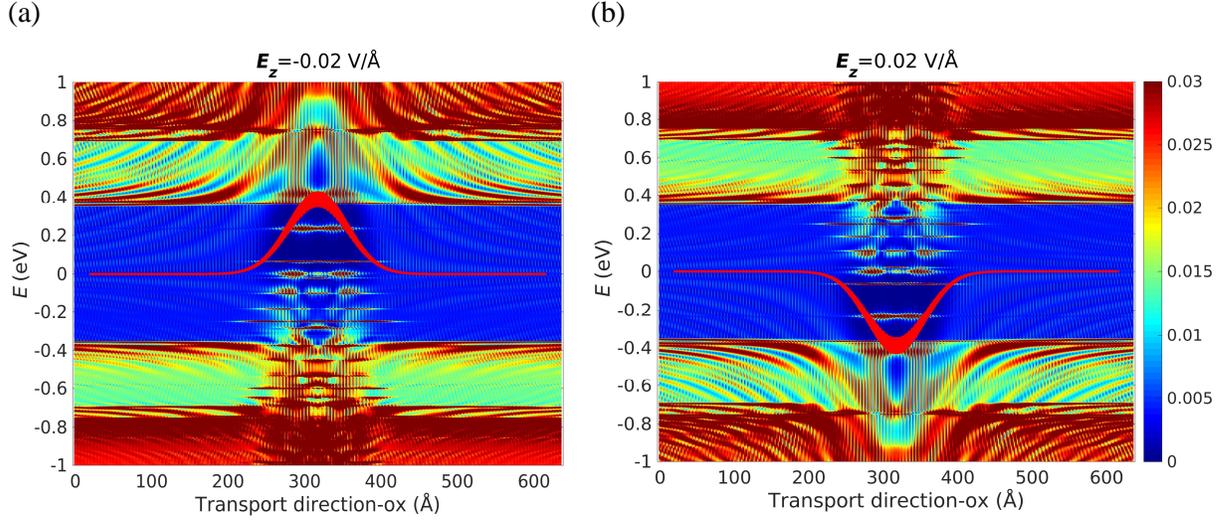

**Fig. 10**: LDOS versus E-ox for opposite fields (a) $\boldsymbol{E}_z$=-0.02 V/Å and (b) $\boldsymbol{E}_z$=0.02 V/Å. Red lines are electrostatic potential profiles induced by the electric fields. Here $M = 41$ ($W_R \approx 49.19$ Å), $N = 150$ ($L \approx 637.58$ Å) and the Gaussian bump $h_G = 21.36$ Å, $\sigma = 39.01$ Å. In both panels, LDOS is in arb. unit.

To better understand the effect of opposite vertical electric fields, we plotted in Fig. 10 the LDOS as a function of energy and the transport direction. The red lines in Fig. 10 display the electrostatic potential at each lattice site along the transport direction. As can be seen in Fig. 10(a), under the effect of the -$\boldsymbol{E}_z$ field, a barrier potential (red line) is established and it causes additional scattering in the region of positive energies. As a consequence, the transmission in the positive energy region drops, as indicated by the violet curve in Fig. 9(b). Besides, this potential profile shifts up states below the barrier, leading to an enhancement of the LDOS in the negative energy region, in particular at high energies in this region (Fig. 10(a)). This explains the enhancement of the electron transport in the negative energy region, as indicated by the violet arrow in Fig. 9(b). It is worth noting that also due to the presence of the potential barrier, there are some strong confinement states within the barrier as seen in Fig. 10(a) and it leads to additional sharp peaks in the transmission near the original two central peaks around $E = 0$ as seen on the violet curve in Fig. 9(b). When the field has the opposite direction (+$\boldsymbol{E}_z$), a quantum well is formed, as shown by the red curve in Fig. 10(b), and the phenomenon is reverse compared to the case of the field -$E_z$.

Similar results were also observed for other groups of armchair ribbons $M = 3p + 1, 3p$.

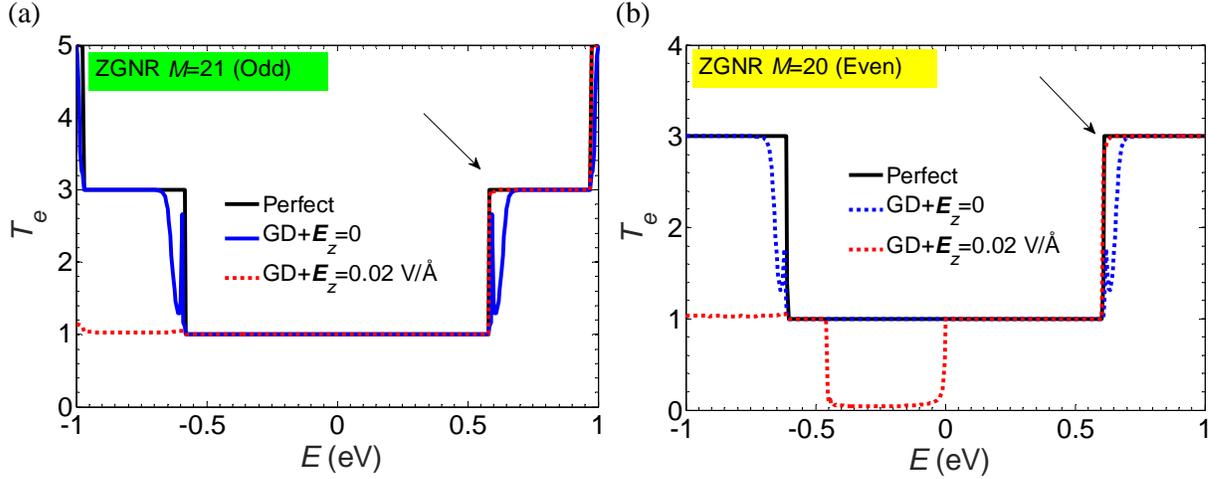

**Fig. 11**: Transmission of ZGNRs without and with a vertical field. (a) Results for odd $M = 21$ ($W_R \approx 43.31$ Å), $N = 150$ ($L \approx 367.69$ Å). (b) Results for even $M = 20$ ($W_R \approx 41.18$ Å), $N = 150$ ($L \approx 367.69$ Å). Both ZGNRs have Gaussian deformation (GD) with the same size $h_G = 25.28$ Å, $\sigma = 46.15$ Å.

We also considered the effect of a vertical electric field on the electronic properties of deformed zigzag ribbons.

Similar to the case of armchair ribbons, an enhancement of the electron transmission was also observed in the conduction region when applying a +$E_z$ field (red line) compared to the case without the field (blue line) as indicated by arrows in Figs. 11(a) and 11(b). And we also observed an inverse effect for the field -$E_z$.

Interestingly, a transport gap in the zigzag structure appears with an even number of chain lines $M$ as shown from the red line in Fig. 11(b). But such a result is not obtained for the odd $M$ zigzag ribbon in Fig. 11(a). In fact, this even-odd effect originates from the well-known parity effect of wave functions in ZGNRs in which the electron transmission is blocked if the right (left) going states $+\vec{k}$ ($-\vec{k}$) of different channels at the same energy level have a different parity[54,55]. The potential induced by the external field leads to a shift of the energy bands in the active region, which results in opposite parities of wave functions and causes a drop of transmission in even *M* zigzag ribbons[54,55].

### C. Edge roughness in Gaussian deformed ribbons

In fabricated ribbons, the edges are commonly not perfect in particular in ribbons made by top-down techniques[6]. It has been also demonstrated that edge roughness strongly impacts on the electronic properties of ribbons[10,15]. In this section, we examine the variation of the electron transport of deformed ribbons in the presence of edge roughness.

To generate edge roughness in a ribbon, $N_{vac}$ atoms were randomly removed from edges. It is noting that the random process can remove atoms in the second line or even in other internal lines from the edges if some border atoms were removed in previous random steps. The level of edge roughness can be defined by the probability to remove atoms at the two edges $P_{ER} = N_{vac}/(4N_A)$. The coefficient 4 is because each unit cell of the perfect structure has 4 atoms at the two edges.

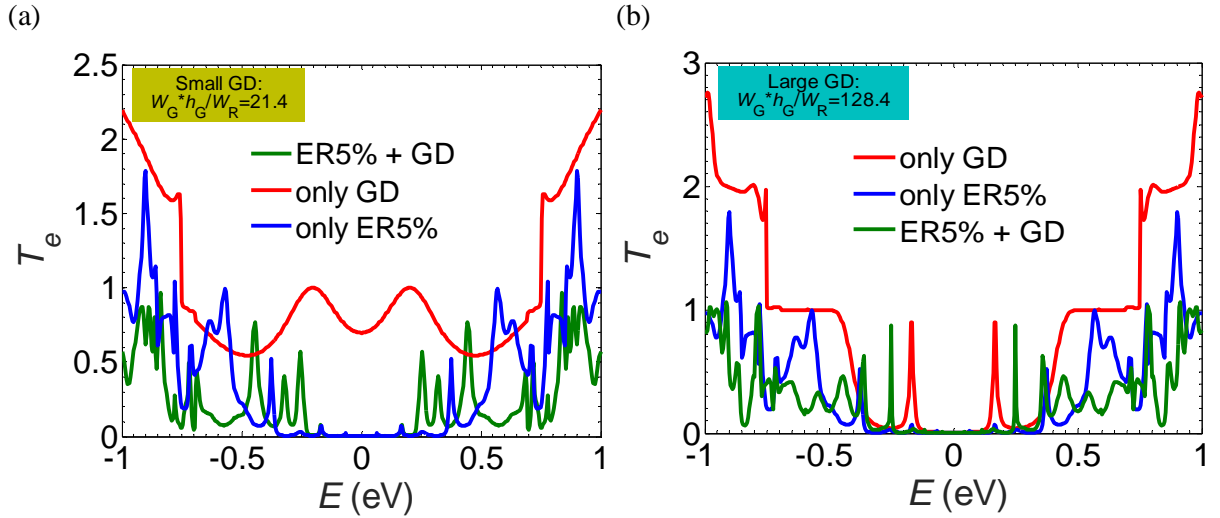

**Fig. 12**: Transmission of a AGNR $M = 20$ ($W_R \approx 23.37$ Å), $N = 60$ ($L \approx 254.18$ Å). Individual and mutual effects of Gaussian deformation (GD) and edge roughness (ER) are considered. The structure with (a) small Gaussian bump $h_G = 6.76$ Å, $\sigma = 12.33$ Å, (b) large Gaussian bump $h_G = 16.55$ Å, $\sigma = 30.21$ Å. In both panels, 5% of ER was considered.

Submitted to Arxiv on 08 May 2020The electron transmission in different structures is shown in Fig. 12: one with a Gaussian bump, one with edge roughness and a structure with both Gaussian deformation and edge roughness. Two cases of Gaussian deformation where a small (Fig. 12(a)) and a large (Fig. 12(b)) Gaussian bumps were considered. In both cases, 5% of edge roughness was considered.

It is worth mentioning that the detailed profile of the transmission of rough ribbons depends on the specific edge-roughness configuration that is stochastic. Thus the transmission should be averaged over many edge configurations. However, as the overall behavior is the same, the results of only one configuration are shown here.

In both panels of Fig. 12, it can be observed that the edge disorder (blue lines) suppresses the electronic transmission more strongly than the Gaussian deformation (red lines). When these two effects are combined, the obtained transmission is dominated by the edge-roughness effect (see green lines). Interestingly, the mutual effect leads to a stronger reduction of the transmission at high energy regions. In contrast, the transport of electrons in the low energy region near $E = 0$ is better than in the case where edge roughness is included, with additional transmission peaks appearing near the zero-energy point (see green lines). Such a phenomenon is due to the formation of some strong Anderson-type localized states at the locations of edge defects[14].

It should be mentioned that similar results were observed for other groups $3p$, $3p + 1$ of armchair and also zigzag ribbons (not shown).

## IV. Conclusion

We have studied the electron transport properties in graphene nanoribbons with Gaussian deformation. Both small and large Gaussian bumps with respect to the size of studied ribbons have been considered. It has shown that Gaussian deformation strongly modifies the electronic properties of all types of ribbon structures. It leads to a strong reduction of electron transmission in high energy regions. In the low energy region and at the first plateau of transmission in semi-metallic armchair ribbons, the transmission is just weakly affected by small Gaussian deformations, however, it drops significantly in the presence of large Gaussian bumps. Besides, the electron transmission can be reduced by 97% in structures exhibiting a sufficiently high ratio $h_G W_G / W_R$ considering the shape of the Gaussian bump over the size of the ribbon. Regarding semi-conducting ribbons, the gap is enlarged when large deformations are applied. The origin of the opening or the widening of the transport gap in large deformed armchair structures has been correlated with the hopping energy profile, i.e., a formation of a three-zone behavior in the hopping profile along the transport direction. Similar electronic characteristics have been observed in the Gaussian folded ribbons. No transport gap is found in deformed zigzag ribbons due to the strong localization of edge states at the energy point $E = 0$. Furthermore, when applying a vertical electric field $+E_z$, the presence of a Gaussian bump breaks the mirror symmetry in vertical planes leading to an enhancement of the electron transport in the conduction region and a degradation in the valence zone. The effect is reverse when employing an opposite field $-E_z$. The study also unveils that the electronic behavior of the deformed ribbon including edge roughness is dominated by the characteristics of edge roughness in which strong sharp peaks in the transmission profile character strong Anderson localization.